\documentclass[twocolumn,showpacs,amsmath,amssymb,pre,aps]{revtex4}   
\usepackage{graphicx}
\usepackage{dcolumn}
\usepackage{bm}
\usepackage{color}

\newif\ifgraph

\graphtrue             %

\begin{document}
\title{
Understanding quorum sensing self-organization: \\
Clustering and defect-induced ordering of diffusing particles
}
\draft

\author{Feifei Liu$^{1}$, Vyacheslav~R.~Misko$^{1,2}$, Yunyun Li$^{1}$ and Fabio Marchesoni$^{1}$}
\affiliation{$^1$ MOE Key Laboratory of Advanced Micro-Structured Materials,
School of Physics Science and Engineering,
Tongji University, Shanghai 200092, China}
\affiliation{$^{2}$
Department of Chemical Engineering, Vrije Universiteit Brussel, 1050 Brussels, Belgium}

\date{\today}

\begin{abstract}
Quorum sensing (QS) is known in biology as a form of intercellular communication mediated by signaling molecules called autoinducers.
The QS protocol governs the transition from individual to collective cell behavior once a critical population density is reached.
Using numerical simulations, we investigate how defects influence the QS transition and the structural organization of the resulting colonies. Our model system consists of a mixture of slow (``cold'') and fast (``hot'') diffusing colloidal particles that obey the QS protocol, together with defect particles characterized by a constant diffusivity.
A striking reentrant solidification of QS particles, characterized by long-range order, is induced by hot defects, whereas cold defects give rise to amorphous structures with only short-range order.
These findings deepen our understanding of the QS interaction and provide a mechanism to control the degree of organization in QS systems, with potential applications in robotics, social sciences, and medicine -- for instance, in overcoming antimicrobial resistance.
\end{abstract}

\maketitle

\section{Introduction}

The concept of quorum sensing (QS) first appeared in biology to describe the intercellular communication system in bacteria, which is mediated by secreted chemical signaling molecules called autoinducers~\cite{bassler,whiteley}.
This process enables an entire bacterial population to collectively regulate gene expression and behavior once the cell density reaches a certain threshold.
Through this intercellular communication, the bacterial population can synchronize its behavior and act as a single multicellular ``organism''~\cite{waters}.
QS regulates a variety of processes in bacterial populations, including bioluminescence, symbiosis, virulence gene expression, antimicrobial resistance, and biofilm formation~\cite{QS1,QS2}.

Among these processes, antimicrobial resistance (AMR) is of particular interest to drug developers, as AMR poses a serious threat to public health~\cite{who-amr,cook}.
Targeting QS rather than directly attacking bacteria offers an alternative approach to reducing antibiotic use and thus preventing AMR.
In this context, inhibiting QS to suppress bacterial virulence may provide distinct advantages for AMR, as it places less selective pressure on pathogens and may hinder the development of resistant bacteria.
The ideal QS inhibitor would have a high minimum inhibitory concentration or high minimum bactericidal concentration (i.e., low antibacterial activity) while remaining an effective disruptor of QS signaling.
Alternatively, new QS inhibitors could be used in combination with existing antibiotics to enhance their efficacy.
Indeed, by disrupting bacterial communication and weakening their defenses, lower antibiotic concentrations may be sufficient to achieve the desired effect.

Anticipating the findings of the present work on the role of defects in cluster formation, structure, and transitions, we note that by introducing defects -- particles that do not undergo the QS transition -- one can control the structure of clusters formed by QS particles.
In terms of multicellular ``organisms'' (i.e., clusters of QS particles), gaining control over their internal structure through defects could provide a useful tool to disrupt the coherence of individual cell actions, for example, to suppress their collective response to antibiotics.
This would help prevent AMR, as individual cells could be more easily neutralized by lower concentrations of antibiotics.

Recently, clustering in QS particle systems has been studied using colloidal particles as a model system, motivated by potential applications in robotics and social sciences~\cite{y-f2024}.
A simple colloidal suspension model was proposed, in which individual particles switch their diffusivity from high (hot) to low (cold) when the local concentration of their nearest neighbors exceeds a certain threshold.
This non-reciprocal interaction mechanism effectively models QS behavior.
By tuning the parameters of the adopted QS protocol, numerical simulations have shown that the suspension can exhibit a variety of two-phase (hot and cold) configurations.

In this work, we investigate the role of defects in the formation of QS particle clusters.
Defects can either naturally exist in a system of QS particles (e.g., ``defected'' biological cells that by any reason are insensitive to autoinducers emitted by other cells, or that cannot emit autoinducers themselves and in that way cannot participate in the QS protocol), or these can be added to the system to impact the QS transition and cluster formation.
The latter scenario is in focus in this work as it allows us to control cluster formation and their structure.
An interesting result is related to the expulsion of foreign inclusions (i.e., defects) from clusters formed by QS particles that can be potentially related to AMR as will be explained below.
While for QS particles we adopt the model of
Ref.~\cite{y-f2024} -- i.e., hot (high diffusivity) particles outside clusters and cold (low diffusivity) particles inside clusters -- we additionally introduce defects: particles that do not undergo the QS transition.
These defect particles have a constant diffusivity, either high or low, independent of local particle density.
We find that this system, depending on the relative concentration of defect particles, exhibits intriguing and counterintuitive behavior.
Cold defects affect clusters to a much lesser extent than hot defects.
Indeed, cold defects behave like ordinary QS particles inside clusters, and their impact on cluster formation is rather limited.
These defects mainly reduce the ability of QS particles to form clusters, especially at high defect concentrations, as they deplete the pool of QS particles available for cluster formation.
However, the effect of hot defects is rather unexpected.
Instead of creating disorder and melting the clusters (as one might naturally expect, since QS particles cool down inside clusters and form a ``solid''), hot defects at intermediate concentrations can have the opposite effect: the degree of ordering can increase!
A detailed analysis of cluster structure reveals an amorphous-to-crystalline transition induced by hot defects.
We explain this unusual defect-induced solidification, which seemingly violates the second law of thermodynamics for equilibrium systems.

The paper is organized as follows.
The model is presented in Sec.~2.
The main numerical results on cluster formation in the presence of defects, along with analyses of cluster density, structure, and phase diagrams, are presented in Sec.~3.
Finally, conclusions are formulated in Sec.~4.

\section{Model}

\subsection{Single-particle dynamics}

We consider a system of overdamped colloidal particles, each labeled by index $i$, undergoing regular two-dimensional Brownian motion described by the Langevin equation
\begin{equation}
\label{LE} \dot{\mathbf{r}}_i = \boldsymbol{\xi}_i(t)
\end{equation}
where $\mathbf{r}_i = (x_i, y_i)$ are the coordinates of the particle's center of mass, and $\boldsymbol{\xi}_i(t) = [\xi_{ix}(t), \xi_{iy}(t)]$ is a stationary Gaussian noise with $\langle \xi_{iq}(t) \rangle = 0$ and $\langle \xi_{iq}(t) \xi_{ip}(0) \rangle = 2D_i \delta_{qp} \delta(t)$ for $q, p = x, y$, modeling equilibrium thermal fluctuations in the surrounding fluid. The particle diffusivity $D_i$ is allowed to switch between a higher and a lower value according to the QS protocol detailed below.
The stochastic differential equations (Eq.~\ref{LE}) were numerically integrated using a standard Euler-Maruyama scheme~\cite{Kloeden}.
To ensure numerical stability, the integrations were performed with a sufficiently short time step, $\delta t = 10^{-3}$.

\subsection{Dynamics of colloidal particles}

We numerically simulated a colloidal suspension by placing $N$ identical, independent colloidal particles obeying Eq.~\ref{LE} in a square box of side $L$ with periodic boundary conditions.
In a colloidal suspension, effects due to the finite size of particles cannot be neglected.
We modeled steric interactions either as:
{\it (i)} hard-disk collisions, treating particles as hard disks of radius $r_0$;
or {\it (ii)} short-range pair repulsions using a truncated Lennard-Jones potential~\cite{CWA},
\begin{equation}
\label{CWA}
V_{ij} =
\begin{cases}
4\epsilon\left[\left(\frac{\sigma}{r_{ij}}\right)^{12} - \left(\frac{\sigma}{r_{ij}}\right)^{6} \right] + \epsilon, & \text{if } r_{ij} \leq r_0 \\[6pt]
0, & \text{otherwise}
\end{cases}
\end{equation}
where $r_{ij}$ is the distance between particles $i$ and $j$, $r_0 = 2^{1/6}\sigma$ locates the potential minimum, and $\sigma = 2r_0$ is the effective particle diameter. In our numerical simulations we set $\epsilon=1$.
Further reciprocal interactions and hydrodynamic interactions~\cite{PNAS,Takagi} have been neglected.

\subsection{Quorum sensing protocol}

\begin{figure}[t]
\centering \includegraphics[width=7cm]{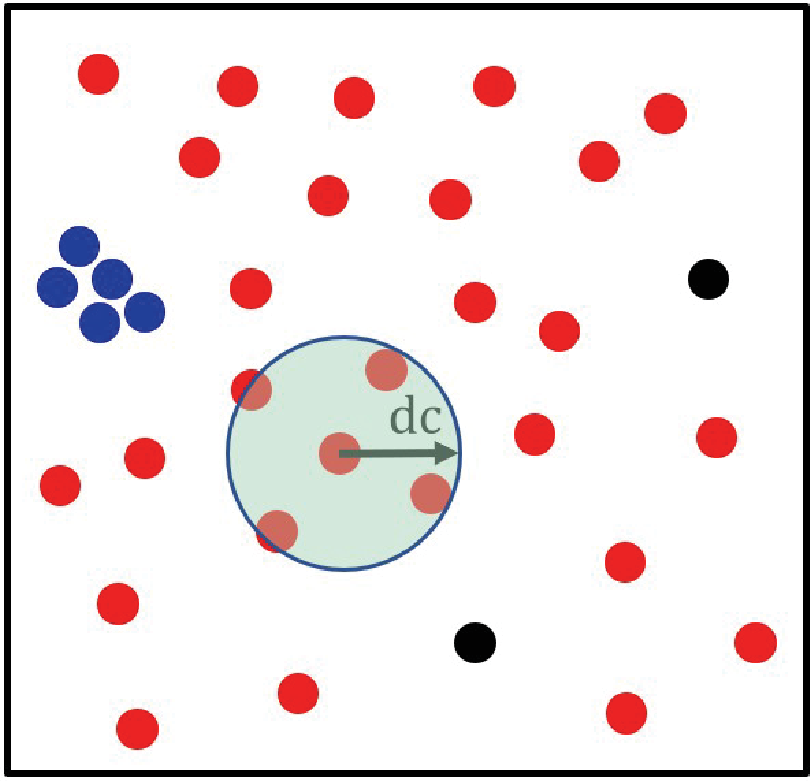}
\caption{Schematic of the QS protocol for the hot-to-cold transition, with $D_{\text{max}} = 1.0$ (red dots) and $D_{\text{min}} = 0.01$ (blue dots). Defect particles are shown as black dots. \label{QS}}
\end{figure}

We assume that the diffusivity of each particle depends on the spatial distribution of its neighbors.
In biological systems, this process is mediated by some form of inter-particle communication (mostly chemical in bacterial colonies~\cite{QS1,QS2}).
It is known that the diffusivity of colloids decreases with increasing density~\cite{Cates2015,Fily}.
The sensing function of particle $i$ can be defined as~\cite{Bauerle2018}
\begin{equation}
\label{Pa}
P_i(d_c) = \sum_{j \in V_i^{d_c}} \frac{1}{2\pi r_{ij}},
\end{equation}
where $V_i^{d_c}$ denotes its perception area of radius $d_c$.
This means that each suspension particle senses the presence of its neighbors in all directions within a restricted horizon $r_{ij} \leq d_c$ (see Fig.~\ref{QS}).

The particle diffusivity is then governed by the following simple QS protocol:
\begin{equation}
\label{QS-e}
D_i =
\begin{cases}
D_{\text{max}}, & P_i(d_c) \leq P_{\text{th}} \\[6pt]
D_{\text{min}}, & P_i(d_c) > P_{\text{th}}
\end{cases}
\end{equation}
where $D_{\text{max}} > D_{\text{min}}$ and the transition threshold is defined as $P_{\text{th}} = \rho_0 L_p$, with $\rho_0 = N/L^2$ denoting the suspension density under uniform spatial distribution.
Note that for a uniform suspension, the sensing function of Eq.~\ref{Pa} is approximately $\bar P(d_c) = \rho_0 d_c$. Clearly, this form of particle interaction is non-reciprocal, since particle $j$ may be perceived by $i$ without itself being affected by the presence of $i$.

\subsection{Tunable model parameters}

A simple rescaling of space and time variables, $x \to x/L$, $y \to y/L$, and $t \to t/\tau$ with $\tau = L^2/D_{\text{min}}$, reveals that the free dynamical parameters of our suspension are $L_p/L$, $d_c/L$, $D_{\text{max}}/D_{\text{min}}$, and $\sigma/L$.
Accordingly, the quantities defining the QS protocol scale as follows: $\rho_0 \to L^2 \rho_0 = N$, $P_{\text{th}} \to L P_{\text{th}} = N(L_p/L)$, and $\bar P(d_c) \to L \bar P(d_c) = N(d_c/L)$.
The choice of the diffusion constants $D_{\text{min}}$ and $D_{\text{max}}$ affects the stationary properties of the suspension only through the ratio $D_{\text{max}}/D_{\text{min}}$.
The particle diameter $\sigma$ enters the simulation output primarily through the definition of the sensing function $P_i(d_c)$ in Eq.~\ref{Pa}, where it serves as a natural cutoff as $r_{ij} \to 0$.
On the other hand, for sensing distances $d_c$ larger than the mean interparticle distance $l_N = \sqrt{L^2/N}$, QS is expected to control suspension diffusivity regardless of short-range steric interactions.
For this reason, in our simulations we fix the suspension parameters $N$ and $L$, the particle size $\sigma = 2r_0$, and therefore the packing fraction $\phi = \pi N (\sigma/L)^2$, while varying the remaining tunable parameters $L_p$, $d_c$, and $D_{\text{max}}/D_{\text{min}}$.

\section{Clustering in a quorum sensing system with defects}

In the absence of defects, a suspension of diffusing particles subject to the QS protocol of Eqs.~(\ref{Pa}) and (\ref{QS-e}) is known to undergo a clustering process \cite{y-f2024}. This occurs over a finite range of the sensing distance $d_c$, above a certain transition threshold that depends on the QS protocol. As particles turn cold, they tend to form small condensation nuclei, which eventually coalesce into a large cluster. Immediately above the transition threshold, this process is critically slow, as the number of transient nuclei decays as $t^{-\alpha}$ with $\alpha \simeq 0.5$. Defects hamper this coalescence process. As shown next, many stable clusters persist even after exceedingly long simulation runs. In the simulations reported here, we set $d_c$ well above the clustering threshold to avoid transient effects and highlight the impact of defects on cluster formation.

\subsection{Cold defects: $D_{\text{defect}} = D_{\text{min}}$}

Let low-diffusivity defects be characterized by the same diffusion coefficient as QS particles after they undergo the QS transition and slow down, i.e., $D_{\text{defect}} = D_{\text{min}}$.
Consequently, these cold defects become indistinguishable from QS particles inside clusters.
Outside clusters, however, cold defects diffuse more slowly than QS particles, which in turn slows down cluster formation compared to the defect-free case.

\begin{figure*}[t]
\centering \includegraphics[width=18cm]{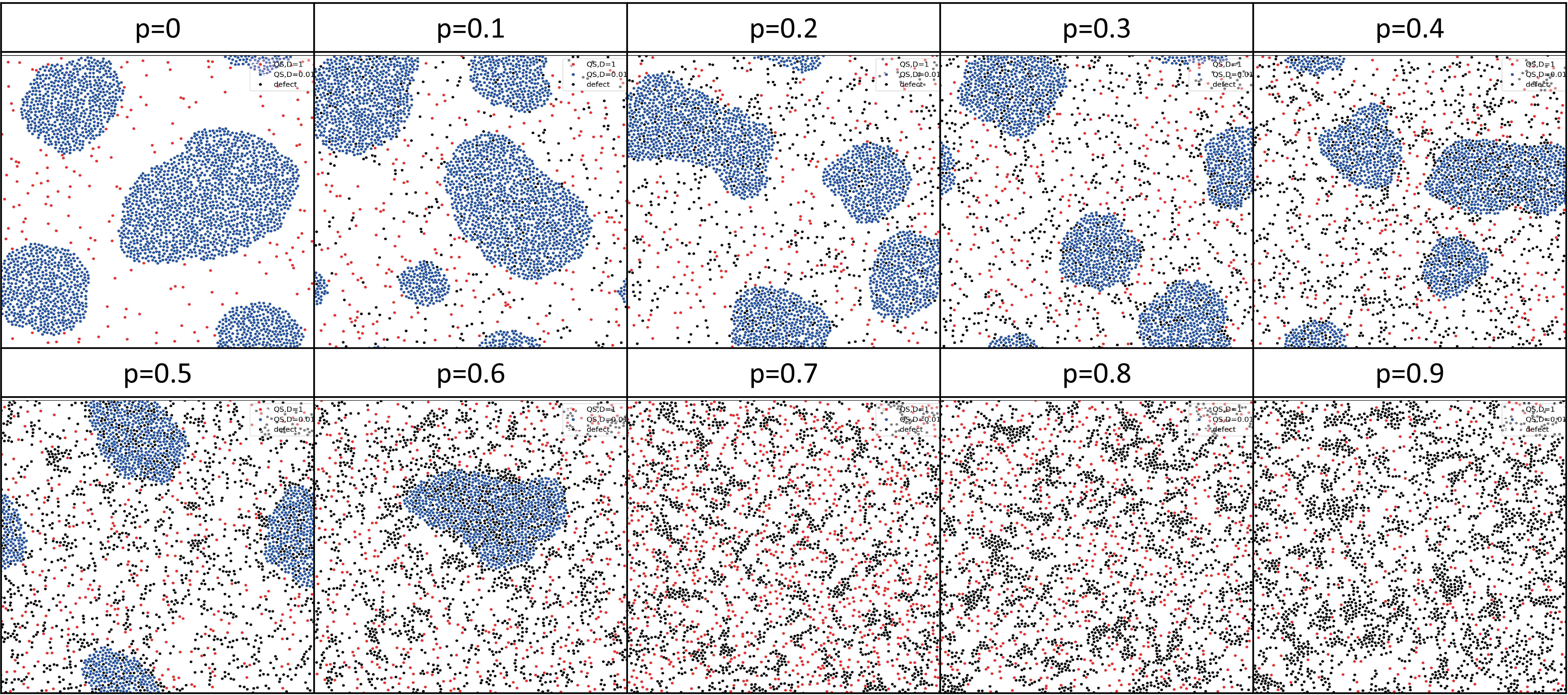}
\caption{Snapshots of particle distributions at $t = 10^5$ for $L_p = 10$, total particle number $N = 3000$, and increasing fractions of cold defect particles (see legends). Other simulation parameters: $D_{\rm min}=0.01$, $D_{\rm max}=1$, $r_0=1$, $\bar \phi=0.15$ and $d_c=L_p=10$. \label{clusters-cold}}
\end{figure*}

The formation of clusters in a system of QS particles in the presence of cold defects at varying concentrations, from $p = 0$ to $p = 0.9$, is shown in Fig.~\ref{clusters-cold}.

Clusters of QS particles are observed for defect concentrations ranging from $p = 0$ to $p = 0.6$, with cluster size and number gradually decreasing as $p$ increases.
This is explained by the gradual reduction in the number of QS particles with increasing defect concentration.
Interestingly, as the defect concentration increases and QS particle clustering is suppressed, we simultaneously observe a different effect: the formation of aggregates (or small clusters) of defect particles. These can already be seen at $p = 0.5$ and $p = 0.6$ while QS particle clusters are still present, and the effect becomes more pronounced at higher defect concentrations ($p = 0.7$ to $p = 0.9$).
We focus on QS particle clustering; the observed formation of small defect clusters (which is clearly weaker than QS clustering) can be regarded as a ``side effect.''
This effect  -- cluster formation in binary mixtures of fast-diffusing and slow-diffusing (i.e., ``hot'' and ``cold'') particles -- has recently attracted considerable interest~\cite{weber,smrek} and its relevance for biological cell sorting highlighted in Ref.~\cite{mccarthy}.
Our calculations show very similar behavior (compare, e.g., our plots for $p = 0.7$ and $p = 0.8$ in Fig.~\ref{clusters-cold} with Fig.~1(a) in Ref.~\cite{weber}).
This effect will be further addressed below.

\begin{figure*}[t]
\centering \includegraphics[width=18cm]{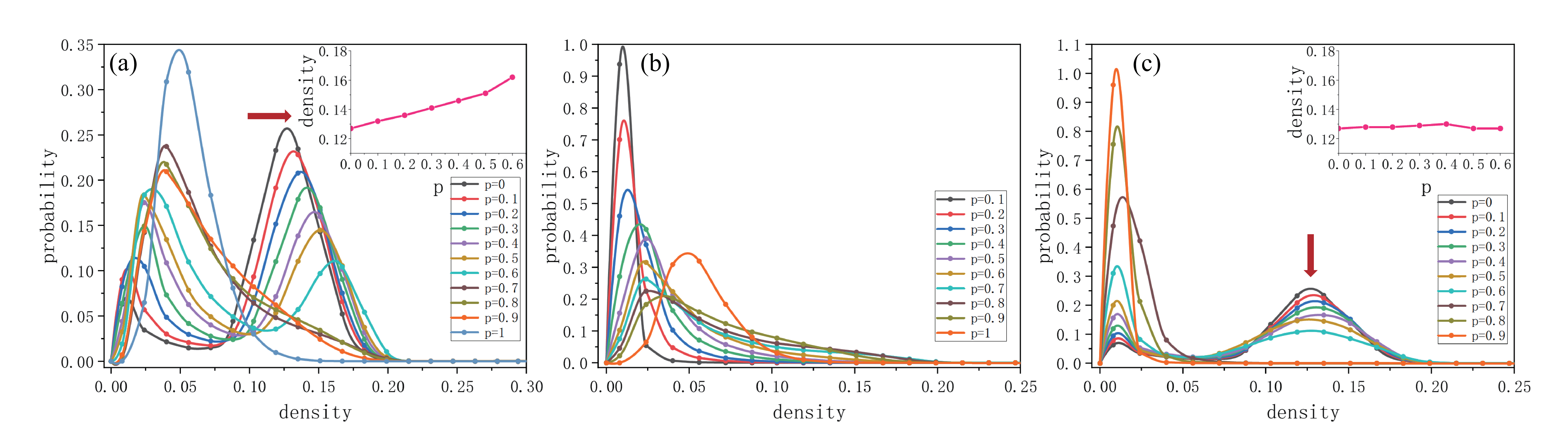}
\caption{Number density distributions for the suspensions of Fig. \ref{clusters-cold} as functions of particle density for (a) all particles (QS and defects), (b) defect particles only, and (c) QS particles only, for increasing fractions of defect particles (see legends).
Insets in (a) and (c) present particle density vs. defect concentration, $p$.
We note that while the density of all particles (QS and defects) increases with $p$, the density of QS particles remains nearly constant.
\label{number-density}}
\end{figure*}

The particle distributions and cluster formation presented in Fig.~\ref{clusters-cold} are further analyzed by calculating the number density function.
The results are presented in Fig.~\ref{number-density} for all particles (including QS particles and defects) (a), defect particles only (b), and QS particles only (c), for varying defect concentrations.

The number density distribution for all particles (Fig.~\ref{number-density}a) shows two maxima: one at low density corresponding to free particles in a gaseous state, and one at high density corresponding to particles in clusters. The position of the high-density peak shifts toward higher densities with increasing defect concentration $p$ for $p = 0$ to $p = 0.6$, i.e., in the range where clusters form.
The corresponding number density distribution for defect particles only (Fig.~\ref{number-density}b) has a single maximum at low densities, which gradually shifts toward higher densities as $p$ increases from $0$ to $0.9$, with a larger shift at $p = 1.0$. Notably, the number density for defect particles is insensitive to the formation of QS clusters.
The number density distribution for QS particles only (Fig.~\ref{number-density}c) also exhibits two maxima (low and high densities), but the high-density peak is suppressed as $p$ increases, indicating that the number of QS particles in clusters decreases with increasing defect concentration.

\begin{figure}[t]
\centering \includegraphics[width=9.5cm]{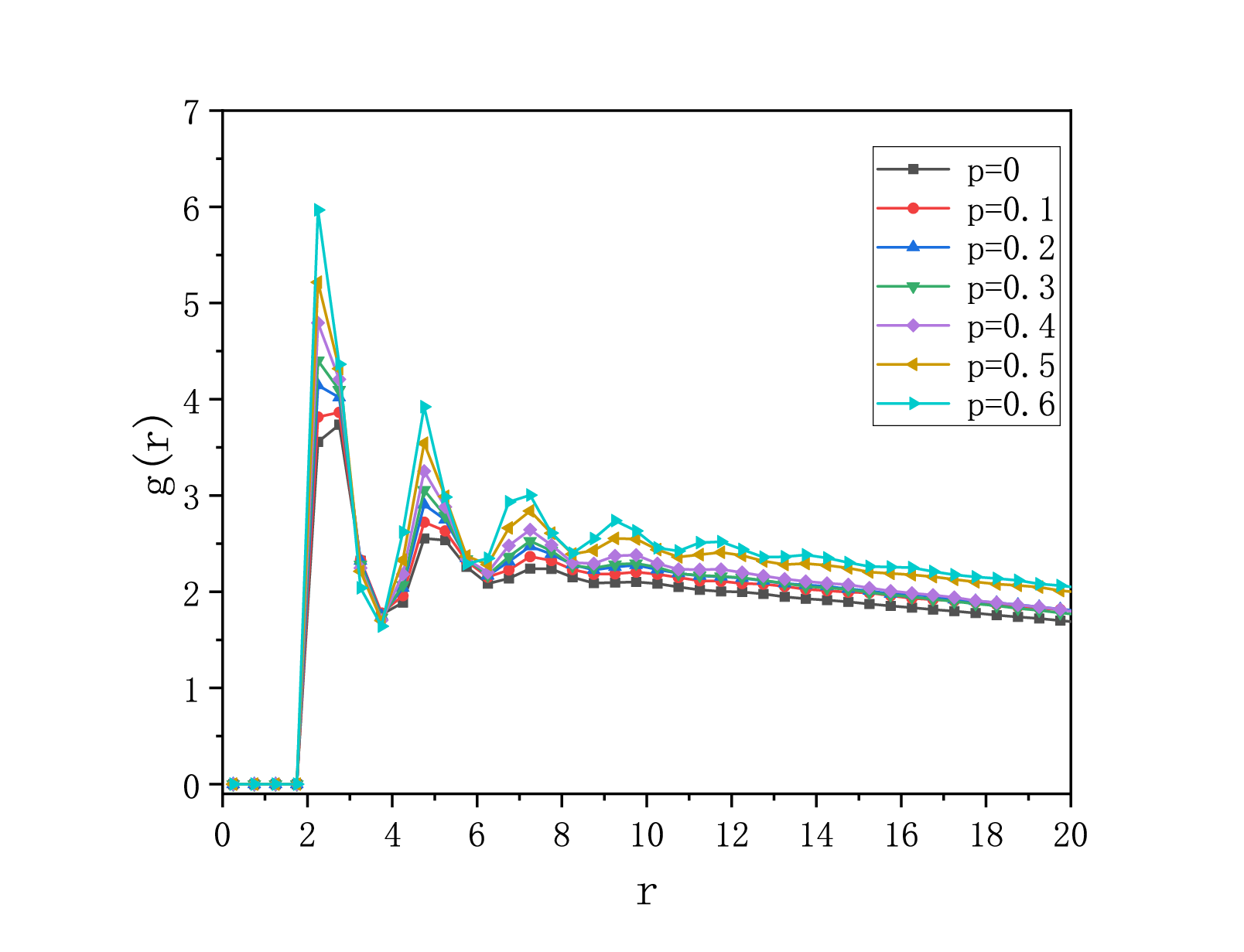}
\caption{Radial distribution function (RDF) for particles in the clusters shown in Fig.~\ref{clusters-cold}, for $p = 0$ to $0.6$. Only the first few peaks are resolved (these become more pronounced for larger $p$), indicating that the particles in the clusters are in an amorphous solid state. \label{rdf-cold}}
\end{figure}

Furthermore, we calculated the radial distribution function (RDF) for particles in clusters.
The results are presented in Fig.~\ref{rdf-cold} for the clusters shown in Fig.~\ref{clusters-cold}, i.e., for $p = 0$ to $0.6$.
Only the first few peaks of the RDF are resolved, and their height becomes more pronounced as $p$ increases (which can be attributed to the role of defects in ordering QS particles within clusters).
This result indicates the absence of long-range order in the cluster structure; the particles are in an amorphous solid state.

\subsection{Hot defects: $D_{\text{defect}} = D_{\text{max}}$}

We now consider a system of QS particles in the presence of ``hot'' defects.
These defects are characterized by the same diffusion coefficient as QS particles outside clusters, i.e., in the low-density limit, $D_{\text{max}}$.
When the sensing function surpasses the threshold, QS particles change their diffusivity from $D_{\text{max}}$ to $D_{\text{min}}$, whereas hot defect particles do not; they remain hot even inside clusters.

One might reasonably expect that the presence of these hot defects would introduce more disorder into the amorphous clusters or even prevent cluster formation altogether.
Indeed, from general thermodynamic considerations, adding hot defects instead of cold defects to a system implies an increase in entropy, i.e., less order.
Let us examine whether these intuitive expectations are indeed correct or whether a different scenario is possible.

\begin{figure*}[t]
\centering \includegraphics[width=18cm]{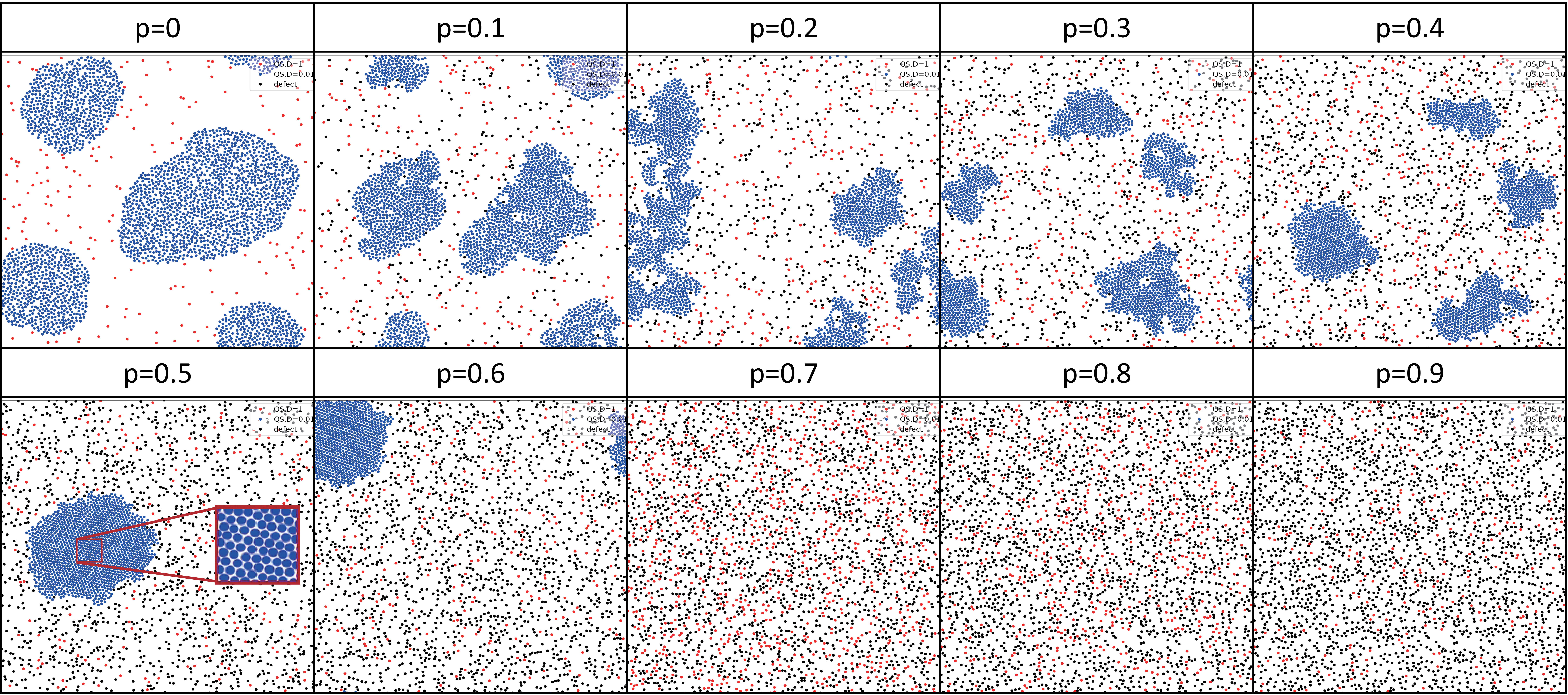}
\caption{Snapshots of particle distributions at $t = 10^5$ for $L_p = 10$, total particle number $N = 3000$, and increasing fractions of hot defect particles (see legends). Other simulation parameters: $D_{\rm min}=0.01$, $D_{\rm max}=1$, $r_0=1$, $\bar \phi=0.15$ and $d_c=L_p=10$.
A zoomed image of particle distribution for $p = 0.5$ shows a perfect hexagonal structure of QS particles in a cluster.
\label{clusters-hot}}
\end{figure*}

The simulation results for cluster formation in a system of QS particles in the presence of hot defects at varying concentrations are presented in Fig.~\ref{clusters-hot}.
As can be seen, clusters form for $p = 0$ to $0.6$ -- the same concentration range as for cold defects (see Fig.~\ref{clusters-cold}).
However, the presence of hot defects results in clusters becoming {\it denser} with increasing defect concentration, contrary to the intuitive expectation that hot defects would suppress clustering.

\begin{figure*}[t]
\centering \includegraphics[width=18cm]{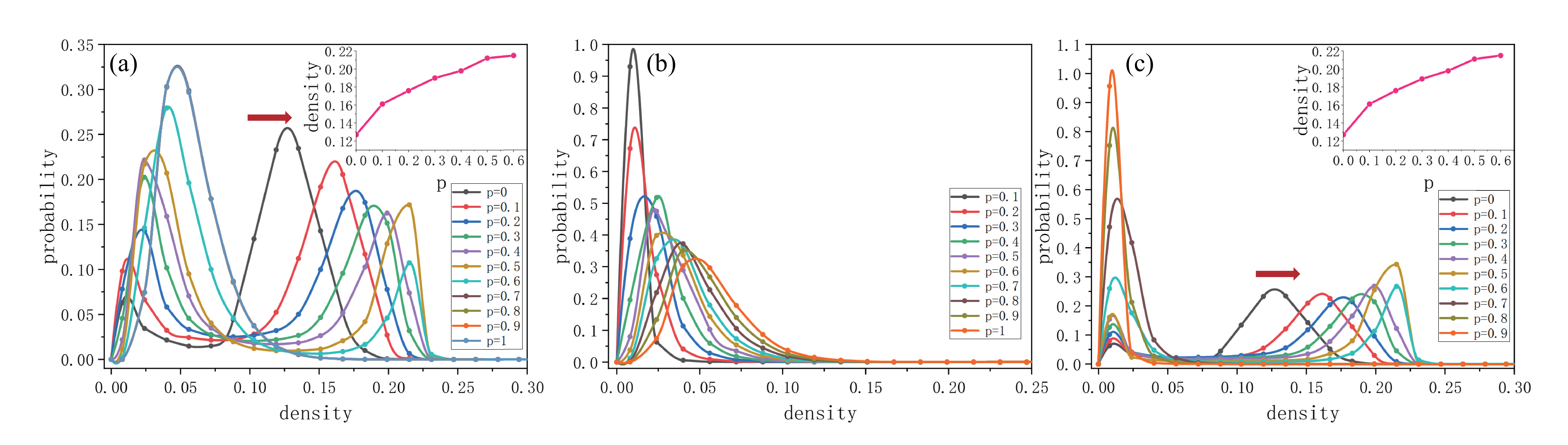}
\caption{Number density distributions for the suspensions of Fig. \ref{clusters-hot} as functions of particle density for hot defects: (a) all particles, (b) defect particles only, and (c) QS particles only, for increasing fractions of defect particles (see legends).
Insets in (a) and (c) present particle density vs. defect concentration, $p$.
The density of QS particles increases with $p$ (cp. Fig.~\ref{number-density}) indicating that the clusters formed by QS particles become denser with increasing $p$.
\label{number-density-hot}}
\end{figure*}

Indeed, the corresponding number density function (Fig.~\ref{number-density-hot}) shows that the density of particles inside clusters increases with defect concentration and is substantially higher than in the case of cold defects (compare with Fig.~\ref{number-density}).

\begin{figure}[t]
\centering \includegraphics[width=9.5cm]{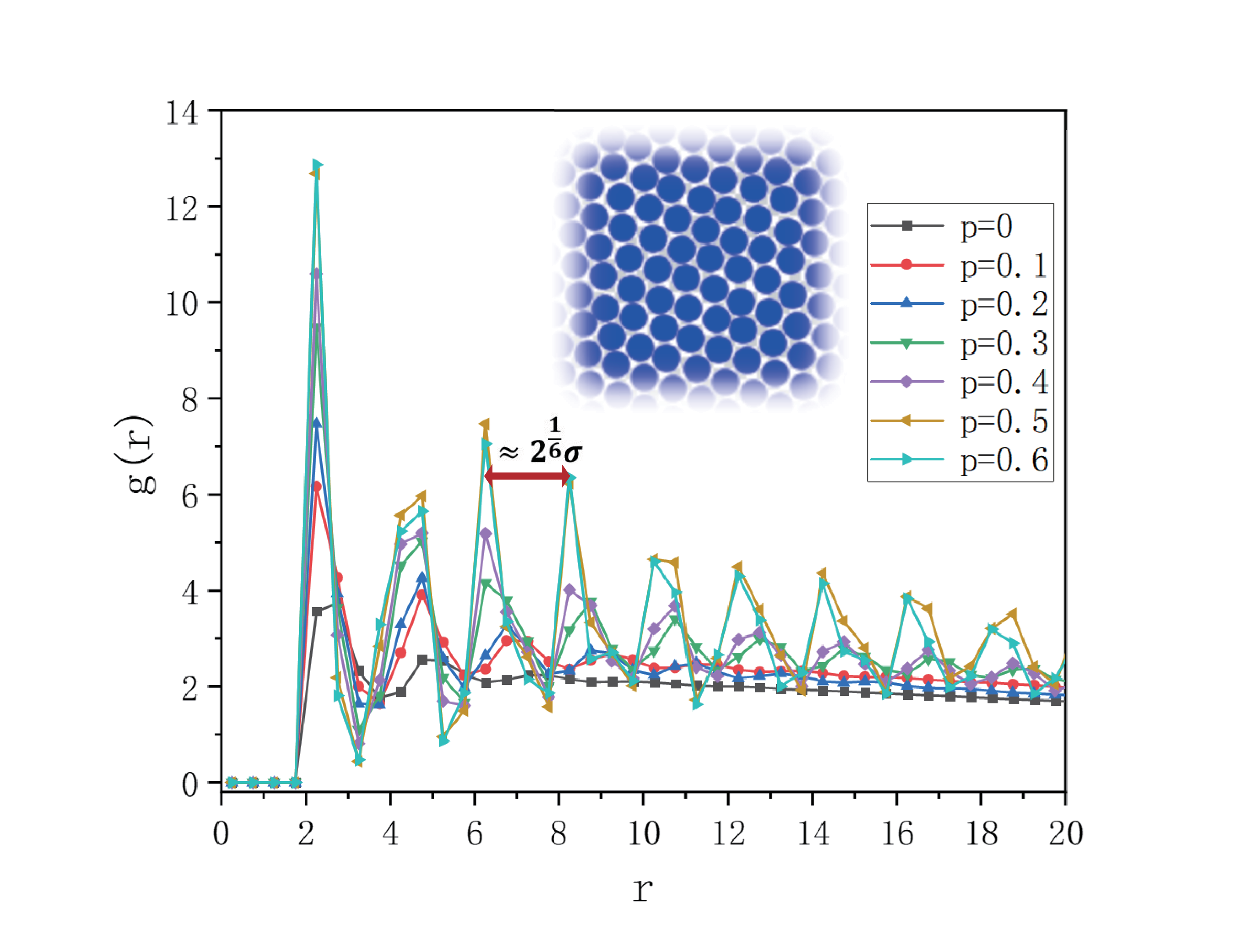}
\caption{Radial distribution function (RDF) for particles in the clusters shown in Fig.~\ref{clusters-hot}, for $p = 0$ to $0.6$. The RDFs exhibit many well-resolved sharp peaks, indicating that the particles are in a crystalline solid state (the inset shows the crystalline structure for $p = 0.6$).
The inter-peak distance of about $2^{1/6}\sigma$ corresponds to densely-packed hexagonal crystalline lattice.
\label{rdf-hot}}
\end{figure}

Furthermore, the RDF for particles in clusters in the presence of hot defects reveals the emergence of striking long-range order.
The corresponding RDFs are presented in Fig.~\ref{rdf-hot} for the clusters shown in Fig.~\ref{clusters-hot}.
The RDF exhibits many peaks for $p \geq 0.3$, and these peaks become sharper and higher for $p = 0.5$ and $0.6$.
Peaks are resolved up to the 9th order (and possibly beyond; only the first nine were calculated, as shown in Fig.~\ref{rdf-hot}).
Thus, the particles in clusters form a perfect hexagonal crystal lattice characterized by long-range order.

Let us attempt to understand this unusual behavior.
Within the paradigm of equilibrium thermodynamics, as mentioned above, defects tend to destroy crystalline order and lead to the well-known phenomenon of defect-induced melting of crystals~\cite{fecht}.
The presence of defects is typically associated with an increase in entropy.

Contrary to this classical scenario, we observe that hot defects in a system of diffusing particles with QS can transform clusters from amorphous into highly ordered, crystalline structures.
In other words, we observe a reentrant behavior: doping, instead of creating disorder, results in ordering (crystallization or ``freezing'').
A similar reentrant behavior was previously observed in driven non-equilibrium systems, known as the ``freezing-by-heating'' transition~\cite{helberg,stanley}, where an increase in fluctuations leads to solidification rather than melting.

However, the system considered here does not involve any external driving, unlike those in Refs.~\cite{helberg,stanley}.
Instead, it incorporates the QS protocol, which allows each particle to ``decide'' its diffusivity at the QS transition point, distinguishing it from a purely diffusive equilibrium system.
The mechanism underlying the observed behavior can be understood as follows.
When QS particles enter a cluster, their diffusion immediately slows down, and they become ``frozen'' in random positions, which results in the formation of amorphous clusters.
When a hot defect enters an amorphous cluster, it remains a fast-diffusing particle (since it does not undergo the QS transition) and explores the surrounding phase space (cf. Refs.~\cite{helberg,stanley}).
As it moves inside the cluster, this fast-diffusing defect effectively pushes particles in its path, squeezing them into a densely packed crystalline structure.
This effect can be thought of as ``shaking the structure from within,'' applying local pressure from the inside due to fast-diffusing defects.
This behavior is similar to that recently observed in a system of passive slowly diffusing particles doped with chemically active particles, which caused crystallization of an initially amorphous passive bead environment~\cite{we-afm}.

Analysis of the evolution of QS clusters reveals that when a fast-diffusing defect moves inside a cluster and squeezes it into a densely packed structure, it creates a void (see Fig.~\ref{clusters-hot} for $p = 0.2$ to $0.4$).
The cluster, however, cannot sustain such voids and self-heals by collapsing them.
During this process, a fast-diffusing defect that was initially trapped (transiently) by the cluster eventually escapes from it (see Fig.~\ref{clusters-hot} for $p = 0.5$ and $0.6$).

Recall that when defects are cold -- i.e., they diffuse with coefficient $D_{\text{min}}$ -- they become indistinguishable from QS particles whose diffusivity also decreases to $D_{\text{min}}$ inside a cluster.
The cluster then remains amorphous, with some QS particles replaced by these slow defects.

Above, we discussed the impact of defects on the structure of clusters of QS particles. It is also interesting, however, to examine the effect of clusters on the defects themselves. As we have shown, fast defects are expelled from clusters during the self-healing process. This expulsion occurs due to the difference in diffusivity between the ``intruders'' (i.e., defects) and the particles of the host subsystem.
Consider for a moment a biological analogy: the host subsystem is a bacterial colony, and the defects are medicine-containing capsules capable of killing the bacteria. According to our findings, if the diffusivity of the capsules differs from that of the host bacterial subsystem, the capsules will be expelled from the colony! In other words, the entire bacterial colony defends itself against the attack of the medicinal capsules, i.e., the medicine becomes ineffective due to the self-organization of the QS bacterial colony (see Ref.~\cite{cook}). This could represent a simple yet plausible mechanism of AMR.
Conversely, if the diffusivity of the medicine-containing capsules is close to that of the host QS bacteria, then the two subsystems will coexist. Consequently, the medicine released from the capsules can then neutralize the bacteria. This suggests a potential strategy for overcoming AMR: match the diffusivity of the therapeutic agents to that of the target bacterial system.
Of course, these are preliminary conclusions drawn from our observations, and this important line of inquiry requires further investigation.

As a final remark, we note that at high defect concentrations ($p = 0.7$ to $0.9$), no aggregation of defect particles is observed in the absence of QS clustering, in contrast to the case of cold defects.
This is because both species -- QS particles (outside clusters) and hot defects -- have the same diffusion coefficient, resulting in a homogeneous distribution.
It is also worth noting that the aggregation mechanism of cold defects observed at high defect concentrations (Fig.~\ref{clusters-cold}) is somewhat similar to the amorphous-to-crystalline transition mechanism presented here, except that fast-diffusing particles push slow-diffusing defects from outside rather than from inside clusters.

\subsection{Phase diagram}

First, we note that defects can be interpreted in different ways in a QS particle system.
Defects can either be counted as particles when calculating the sensing function (Eq.~\ref{Pa}) within the perception area, or not.
All results presented above correspond to the latter case, i.e., defects do not contribute to the sensing function.
This choice is physically reasonable, as defects do not undergo the QS transition, and we should count only those particles that do (i.e., QS particles).
Indeed, if we imagine a bacterial colony as a QS subsystem and add synthetic particles that do not emit autoinducers, these particles would not be included in the sensing function.
On the other hand, we can imagine a situation where defect particles participate in the overall perception mechanism (thus contributing to the sensing function) but do not change their diffusivity as required by the QS protocol.
This could be the case, for example, of bacteria that emit and sense autoinducers but do not properly respond to these stimuli (unlike QS bacteria).
Such defects would then contribute to the sensing function (Eq.~\ref{Pa}).
We calculated all particle distributions (and their evolution) along with the corresponding quantifiers (number density and RDF) also for the ``count'' defects case, for comparison.
These results, however, are not shown because they are quite similar to those presented. We only show the phase diagrams for both cases below. Comparing these phase diagrams provides insight into how the two cases relate to one another.

\begin{figure*}[t]
\centering \includegraphics[width=18.5cm]{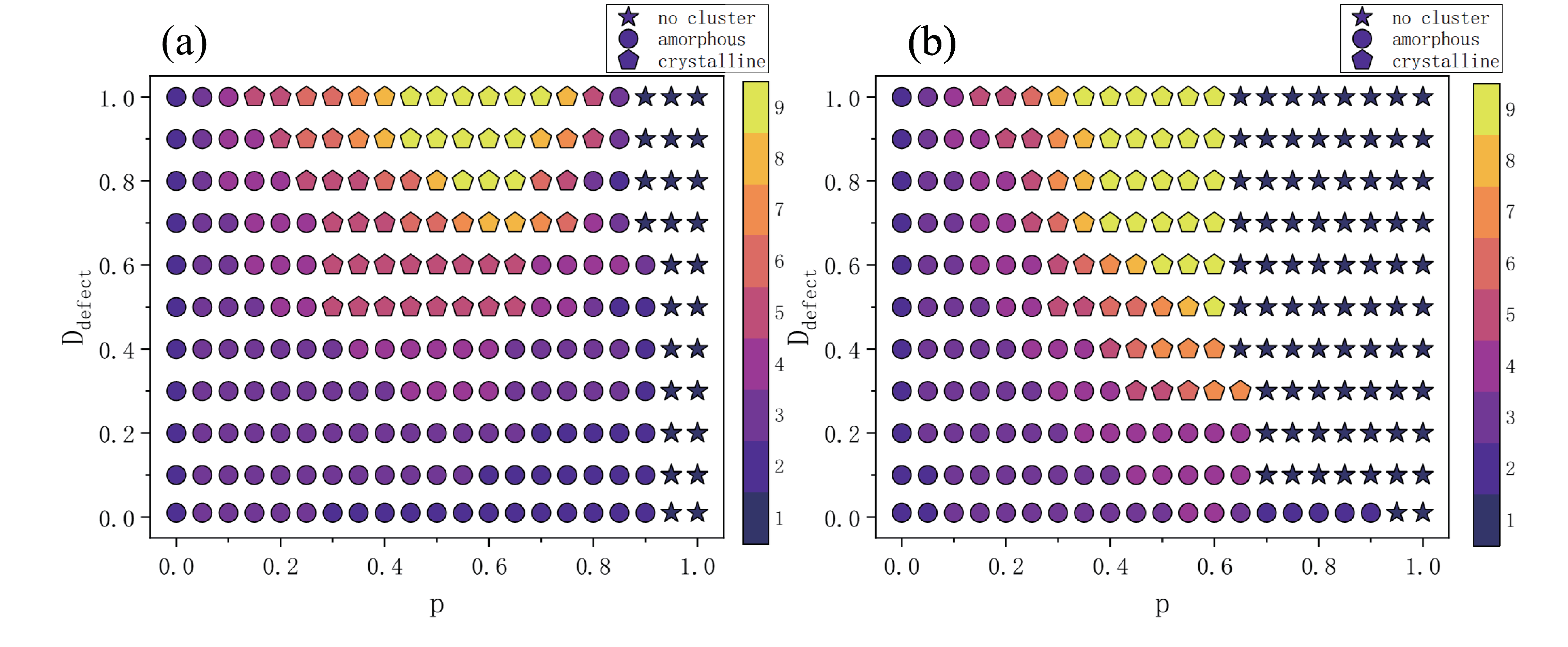}
\caption{Phase diagram in the plane $D_{\rm defect}$ vs.  $p$, showing different morphologies in a system of diffusing QS particles in the presence of defects, for the ``count'' defects (a) and ``no-count'' defects (b) protocols. The shape of the data points denotes the phase of the suspension (see legends), while their color indicates the number of peaks in the relevant RDF (see color code bar). Here, $D_{\rm min}=0.01$, $D_{\rm max}=1$, $r_0=1$, $\bar \phi=0.15$ and $d_c=L_p=10$.
\label{phase-diagram}}
\end{figure*}

Based on the analysis presented above, we can distinguish different aggregation structures that represent true ``phases'' (though not in a strict thermodynamic sense) and construct the phase diagram in the plane ``$D_{\text{defect}}$ vs. $p$'' (Fig.~\ref{phase-diagram}).

\vspace{1cm}

The phase diagram reveals the following phases:
(i) an ``amorphous solid'' for low defect diffusivity and low defect concentration;
(ii) a highly ordered ``crystalline'' phase for high defect diffusivity and intermediate defect concentration; and
(iii) no clustering at high defect concentration.
In addition, an ``inverted'' state can be observed at high defect concentration in the presence of QS particles (see Fig.~\ref{clusters-cold} for $p = 0.9$).
This occurs when defects are slow-diffusing and the QS particle concentration is very low. In that case, fast QS particles cannot undergo the QS transition.
These QS particles remain fast-diffusing and play the role of ``fast-diffusing defects'' (as observed above at low defect concentrations), thus facilitating the formation of clusters of slowly diffusing defects (see Fig.~\ref{clusters-cold}, $p = 0.9$).

Finally, for pure defect particles, no clusters can form, and the particle distribution is homogeneous.
Therefore, an important conclusion is that cluster formation is only possible in binary systems: either (a) fast-diffusing QS particles together with QS particles that have turned to a slow-diffusion regime following the QS protocol, or (b) QS particles together with defects (which also differ in diffusivity). In the latter case, the amorphous-to-crystalline transition can be observed.

We also note that the phase boundary between the amorphous and crystalline phases is smooth, indicating a continuous crossover, whereas the boundary separating the crystalline phase from the no-cluster phase is sharp (Fig.~\ref{phase-diagram}). This implies that amorphous clusters gradually become crystalline with increasing defect concentration, but the highly ordered clusters disappear abruptly.
The sudden suppression of clustering is particularly pronounced at high densities of diffusive ``no-count'' defects.
This abrupt transition can be understood by noting that, for a QS particle to sense the presence of its neighbors, its sensing range $d_c$ must be comparable to the average interparticle distance $l_N = \sqrt{\pi r_0^2 / \bar \phi}$ -- more precisely, $d_c - \sigma \simeq l_N$.
When a fraction $p$ of ``no-count'' defects is introduced, the effective packing fraction of QS particles that contribute to the sum in Eq.~(\ref{Pa}) reduces to $(1-p)\bar \phi$.
Consequently, clustering is not expected for
$p > 1 - (\pi / \bar \phi)[r_0 / (d_c - \sigma)]^2 \simeq 0.67$, regardless of $D_{\text{defect}}$, in reasonably good agreement with the right panel of Fig.~\ref{phase-diagram}.
This mechanism does not apply to ``count'' defects, since they do contribute to the sum in Eq.~(\ref{Pa}). However, for $l_N \gtrsim 2d_c$, i.e., for $p > 0.95$ in Fig.~\ref{phase-diagram}, the nonreciprocal interaction of Eqs.~(\ref{Pa}) and (\ref{QS-e}) between a pair of QS particles surrounded by any sort of defects, becomes strongly suppressed, increasingly so as $D_{\text{defect}}$ approaches zero.
Consequently, in the left panel of Fig.~\ref{phase-diagram} the no-cluster phase shifts to $p$ values much closer to unity, and the transition from crystalline to amorphous to no-cluster becomes smoother and more $D_{\text{defect}}$ sensitive.

The current analysis of the phase diagram,
Fig.~\ref{phase-diagram},
highlights the key role of defects in the phase formation.
Thus, for pure QS particles, clusters form but remain amorphous.
At high defect concentrations, no clusters form at all.
When defects are present at intermediate concentrations, they lead to the formation of a crystalline phase.
The mechanism involves the motion of fast-diffusing defects inside amorphous clusters, which compress these clusters from within, creating voids around the fast-diffusing defect. The cluster then collapses via surface tension (which tends to minimize the cluster's surface area), expelling the defect.
At high defect concentrations (but still not too close to 1), the opposite behavior is observed: low-concentration QS particles do not undergo the QS transition, but these fast-diffusing particles compress the lower-diffusivity defect particles into clusters, analogous to how fast-diffusing defects act inside QS clusters.
When the defect concentration approaches 1, the effect of these QS particles vanishes, and the clusters dissolve, resulting in a homogeneous particle distribution.

\section{Conclusions}

Inspired by quorum sensing (QS) in bacterial colonies, we studied the effect of defects on a system of diffusing particles that undergo a QS transition when the local density within their perception area exceeds a certain threshold.
The diffusivity of QS particles then abruptly changes from a high initial value to a lower value; i.e., hot particles ``cool down.'' We considered two types of defects: cold defects (always low diffusivity) and hot defects (always high diffusivity).

We found that in both cases, defects can assist in the formation of QS particle clusters.
In the case of cold defects, these particles are indistinguishable from QS particles inside clusters. The main difference from the defect-free case is that increasing defect concentration reduces the fraction of QS particles, thereby suppressing cluster formation. The internal structure of the clusters remains amorphous, as hot QS particles arrive randomly at the boundary of the perception area and become frozen at random positions, with slow diffusion insufficient to alter this amorphous structure.

In contrast, the effect of hot defects on QS cluster formation is fundamentally different.
Instead of weakening the cluster structure (or even melting it, as one might expect from a thermodynamic perspective), increasing concentrations of hot defects induce an amorphous-to-crystalline transition, creating long-range order within the QS clusters.
The calculated radial distribution function  reveals many sharp, well-resolved peaks, indicating a high degree of ordering.
Thus, we have demonstrated a highly unusual effect induced by fast defects in a diffusing particle system with QS: the creation of ordered crystalline-like structures.
This stands in sharp contrast to well-known effects associated with adding fast defects to particle systems, such as melting of colloidal crystals~\cite{fecht} or stirring~\cite{sen,nanoscale2020}.

Furthermore, we presented a phase diagram in the plane  $D_{\rm defect}$ vs.  $p$, showing different morphologies including amorphous clusters, crystalline clusters, and no-cluster states.

Finally, our results may provide a deeper understanding of the behavior of various biological systems, such as bacterial colonies or artificial systems, such as microrobots.
Our findings offer new insight into the ``code'' of QS and provide a tool for controlling the degree of interaction (and intercommunication) among QS particles in a group by introducing different types of defects (cold or hot).
As an example of the collective response of a QS system, we demonstrated the expulsion of defects (e.g., capsules with medicine) that differ in diffusivity from that of the host QS particles (e.g., bacteria) thus providing a possible simple interpretation of antimicrobial resistance, and offering a solution for overcoming AMR: match the diffusivity of the therapeutic agents to that of the target bacterial system.
This task requires and stimulates further deeper study.
We therefore believe these results could be of interest in unlocking the QS code responsible for the collective response of bacteria to external stimuli such as antibiotics, and could thus potentially contribute to research on antimicrobial resistance.

\smallskip

{\it Acknowledgments.---} Y.L. is supported by the NSF China under grant No. 12375037,
and F.M. also by NSF China under grant No. 12350710786.

\end{document}